\documentclass[12pt]{article}
\usepackage{graphics,graphicx}
\usepackage{amssymb,epsfig,amsmath,euscript,array}
\usepackage{cite}
\usepackage{axodraw}
\usepackage{pstricks}
\usepackage{color}

\makeatletter
\@addtoreset{equation}{section}
\makeatother

\newcounter{multieqs}

\newcommand{\be}{\begin{equation}}
\newcommand{\ee}{\end{equation}}

\newcommand{\bm}[1]{\mbox{\boldmath $#1$}}

\newcommand{\kslash}{k \!\!\! / }

\newcommand{\lslash}{l \!\! / }
\newcommand{\Pslash}{P \!\!\!\! / }

\newcommand{\islash}{i \!\!\! / }
\newcommand{\jslash}{j \!\!\! / }
\newcommand{\aslash}{a \!\!\! / }
\newcommand{\bslash}{{b \hspace{-6pt} \slash} }

\newcommand{\onslash}{1 \!\!\! / }
\newcommand{\twslash}{2 \!\!\!/ }
\newcommand{\thslash}{3 \!\!\!/ }
\newcommand{\foslash}{4 \!\!\! / }
\newcommand{\fislash}{5 \!\!\! / }

\newcommand{\mslash}{m \!\!\! / }

\def\bd{\begin{document}}
\def\ed{\end{document}}
\def\nn{\nonumber}
\def\bea{\begin{eqnarray}}
\def\eea{\end{eqnarray}}

\def\ab{(ijab)}
\def\ba{(ijba)}
\def\ijab{{\tr}_{+}(\islash\, \jslash\, \aslash \, \bslash)}
\def\ijba{{\tr}_{+}(\islash\, \jslash\, \bslash \, \aslash)}
\def\ijaP{{\tr}_{+}(\islash\, \jslash\, \aslash \, \Pslash)}
\def\ijPLa{{\tr}_{+}(\islash\, \jslash\, \Pslash_L \, \aslash)}
\def\ijaPL{{\tr}_{+}(\islash\, \jslash\, \aslash \, \Pslash_L)}
\def\ijPLza{{\tr}_{+}(\islash\, \jslash\, \Pslash_{L;z} \, \aslash)}
\def\ijaPLz{{\tr}_{+}(\islash\, \jslash\, \aslash \, \Pslash_{L;z})}
\def\ijPa{{\tr}_{+}(\islash\, \jslash\, \Pslash \, \aslash)}
\def\iaPb{{\tr}_{+}(\islash\, \aslash\, \Pslash \, \bslash)}
\def\ibPa{{\tr}_{+}(\islash\, \bslash\, \Pslash \, \aslash)}
\def\ijPmu{{\tr}_{+}(\islash\, \jslash\, \Pslash \, \mu)}
\def\ibmuP{{\tr}_{+}(\islash\, \bslash\, \mu \, \Pslash)}
\def\ibmua{{\tr}_{+}(\islash\, \bslash\, \mu \, \aslash)}
\def\iamub{{\tr}_{+}(\islash\, \aslash\, \mu \, \bslash)}
\def\jaPb{{\tr}_{+}(\jslash\, \aslash\, \Pslash \, \bslash)}
\def\ijmuP{{\tr}_{+}(\islash\, \jslash\, \mu \, \Pslash)}
\def\ijmum{{\tr}_{+}(\islash\, \jslash\, \mu \, \mslash)}
\def\ijmmu{{\tr}_{+}(\islash\, \jslash\, \mslash \, \mu)}
\def\ijmP{{\tr}_{+}(\islash\, \jslash\, \mslash \, \Pslash)}
\def\iabP{{\tr}_{+}(\islash\, \aslash\, \bslash \, \Pslash)}
\def\ijbP{{\tr}_{+}(\islash\, \jslash\, \bslash \, \Pslash)}
\def\jbPa{{\tr}_{+}(\jslash\, \bslash\, \Pslash \, \aslash)}
\def\ijPb{{\tr}_{+}(\islash\, \jslash\, \Pslash \, \bslash)}
\def\jbmua{{\tr}_{+}(\jslash\, \bslash\, \mu \, \aslash)}

\def\loablt{ {\tr}_{+}(\lslash_1\, \aslash \, \bslash\, \lslash_2)}

\def\ijlolt{{\tr}_{+}(\islash\, \jslash\, \lslash_1 \, \lslash_2)}
\def\ijltlo{{\tr}_{+}(\islash\, \jslash\, \lslash_2 \, \lslash_1)}
\def\ibloa{{\tr}_{+}(\islash\, \bslash\, \lslash_1 \, \aslash)}
\def\jaltb{{\tr}_{+}(\jslash\, \aslash\, \lslash_2 \, \bslash)}
\def\ialtb{{\tr}_{+}(\islash\, \aslash\, \lslash_2 \, \bslash)}
\def\bltloa{{\tr}_{+}(\bslash\, \lslash_2\, \lslash_1 \, \aslash)}
\def\jbloa{{\tr}_{+}(\jslash\, \bslash\, \lslash_1 \, \aslash)}
\def\ibPb{{\tr}_{+}(\islash\, \bslash\, \Pslash \, \bslash)}
\def\ijltb{{\tr}_{+}(\islash\, \jslash\, \lslash_2 \, \bslash)}

\def\ijloa{{\tr}_{+}(\islash\, \jslash\,  \lslash_1 \, \aslash)}
\def\ijblt{{\tr}_{+}(\islash\, \jslash\,  \bslash \, \lslash_2)}

\def\jakb{{\tr}_{+}(\jslash\, \aslash\, \kslash \, \bslash)}
\def\iakb{{\tr}_{+}(\islash\, \aslash\, \kslash \, \bslash)}

\def\tofo{{\tr}_{+}(\onslash\, \thslash\, \twslash \, \foslash)}
\def\foto{{\tr}_{+}(\onslash\, \thslash\, \foslash \, \twslash)}
\def\tofi{{\tr}_{+}(\onslash\, \thslash\, \twslash \, \fislash)}
\def\fito{{\tr}_{+}(\onslash\, \thslash\, \fislash \, \twslash)}

\def\lrangle#1#2{\langle #1\,#2\rangle}

\def\Li{{$\rm Li}_2$}
\def\eps{\epsilon}
\def\epsuv{{\epsilon_{\rm \mbox{\tiny UV}}}}
\let\bm=\bibitem
\let\la=\label

\def\npb#1#2#3{Nucl. Phys. {\bf{B#1}} #3 (#2)}
\def\plb#1#2#3{Phys. Lett. {\bf{#1B}} #3 (#2)}
\def\prl#1#2#3{Phys. Rev. Lett. {\bf{#1}} #3 (#2)}
\def\prd#1#2#3{Phys. Rev. {D \bf{#1}} #3 (#2)}
\def\cmp#1#2#3{Comm. Math. Phys. {\bf{#1}} #3 (#2)}
\def\cqg#1#2#3{Class. Quantum Grav. {\bf{#1}} #3 (#2)}
\def\nppsa#1#2#3{Nucl. Phys. B (Proc. Suppl.) {\bf{#1A}}#3 (#2)}
\def\ap#1#2#3{Ann. of Phys. {\bf{#1}} #3 (#2)}
\def\ijmp#1#2#3{Int. J. Mod. Phys. {\bf{A#1}} #3 (#2)}
\def\rmp#1#2#3{Rev. Mod. Phys. {\bf{#1}} #3 (#2)}
\def\mpla#1#2#3{Mod. Phys. Lett. {\bf A#1} #3 (#2)}
\def\jhep#1#2#3{J. High Energy Phys. {\bf #1} #3 (#2)}
\def\atmp#1#2#3{Adv. Theor. Math. Phys. {\bf #1} #3 (#2)}
\newcommand{\EQ}[1]{\begin{equation} #1 \end{equation}}
\newcommand{\AL}[1]{\begin{subequations}\begin{align} #1 \end{align}\end{subequations}}
\newcommand{\SP}[1]{\begin{equation}\begin{split} #1 \end{split}\end{equation}}
\newcommand{\ALAT}[2]{\begin{subequations}\begin{alignat}{#1} #2 \end{alignat}
                        \end{subequations}}
\def\beqa{\begin{eqnarray}}
\def\eeqa{\end{eqnarray}}
\def\beq{\begin{equation}}
\def\eeq{\end{equation}}
\def\sst{\scriptscriptstyle}
\def\thetabar{\bar\theta}
\def\Tr{{\rm Tr}}
\def\one{\mbox{1 \kern-.59em {\rm l}}}
 \def\Nh{\hat{N}}

\newcommand{\half}{{\textstyle {1 \over 2}}}

\def\a{\alpha}      \def\da{{\dot\alpha}}
\def\b{\beta}       \def\db{{\dot\beta}}
\def\c{\gamma}  \def\G{\Gamma}  \def\cdt{\dot\gamma}
\def\d{\delta}  \def\D{\Delta}  \def\ddt{\dot\delta}
\def\e{\epsilon}        \def\vare{\varepsilon}
\def\f{\phi}    \def\F{\Phi}    \def\vvf{\f}
\def\h{\eta}
\def\k{\kappa}
\def\l{\lambda} \def\L{\Lambda}
\def\m{\mu} \def\n{\nu}
\def\o{\omega}
\def\p{\pi} \def\P{\Pi}
\def\r{\rho}
\def\s{\sigma}  \def\S{\Sigma}
\def\t{\tau}
\def\th{\theta} \def\Th{\Theta} \def\vth{\vartheta}
\def\X{\Xeta}
\def\z{\zeta}
\def\de{\partial}

\def\cA{{\cal A}} \def\cB{{\cal B}} \def\cC{{\cal C}}
\def\cD{{\cal D}} \def\cE{{\cal E}} \def\cF{{\cal F}}
\def\cG{{\cal G}} \def\cH{{\cal H}} \def\cI{{\cal I}}
\def\cJ{{\cal J}} \def\cK{{\cal K}} \def\cL{{\cal L}}
\def\cM{{\cal M}} \def\cN{{\cal N}} \def\cO{{\cal O}}
\def\cP{{\cal P}} \def\cQ{{\cal Q}} \def\cR{{\cal R}}
\def\cS{{\cal S}} \def\cT{{\cal T}} \def\cU{{\cal U}}
\def\cV{{\cal V}} \def\cW{{\cal W}} \def\cX{{\cal X}}
\def\cY{{\cal Y}} \def\cZ{{\cal Z}}

\def\ua{\underline{\alpha}}
\def\ub{\underline{\phantom{\alpha}}\!\!\!\beta}
\def\uc{\underline{\phantom{\alpha}}\!\!\!\gamma}
\def\um{\underline{\mu}}
\def\ud{\underline\delta}
\def\ue{\underline\epsilon}
\def\una{\underline a}\def\unA{\underline A}
\def\unb{\underline b}\def\unB{\underline B}
\def\unc{\underline c}\def\unC{\underline C}
\def\und{\underline d}\def\unD{\underline D}
\def\une{\underline e}\def\unE{\underline E}
\def\unf{\underline{\phantom{e}}\!\!\!\! f}\def\unF{\underline F}
\def\unm{\underline m}\def\unM{\underline M}
\def\unn{\underline n}\def\unN{\underline N}
\def\unp{\underline{\phantom{a}}\!\!\! p}\def\unP{\underline P}
\def\unq{\underline{\phantom{a}}\!\!\! q}
\def\unQ{\underline{\phantom{A}}\!\!\!\! Q}
\def\unH{\underline{H}}

\def\As {{A \hspace{-6.4pt} \slash}\;}
\def\bs {{b \hspace{-6.4pt} \slash}\;}
\def\Ds {{D \hspace{-6.4pt} \slash}\;}
\def\ds {{\del \hspace{-6.4pt} \slash}\;}
\def\ss {{\s \hspace{-6.4pt} \slash}\;}
\def\ks {{ k \hspace{-6.4pt} \slash}\;}
\def\ps {{p \hspace{-6.4pt} \slash}\;}
\def\pas {{{p_1} \hspace{-6.4pt} \slash}\;}
\def\pbs {{{p_2} \hspace{-6.4pt} \slash}\;}
\def\Ps {{P \hspace{-6.4pt} \slash}\;}
\def\Qs {{Q \hspace{-6.4pt} \slash}\;}

\def\Fh{\hat{F}}
\def\Vh{\hat{V}}
\def\Xh{\hat{X}}
\def\ah{\hat{a}}
\def\xh{\hat{x}}
\def\yh{\hat{y}}
\def\ph{\hat{p}}
\def\xih{\hat{\xi}}
\def\psit{\tilde{\psi}}
\def\Psit{\tilde{\Psi}}
\def\tht{\tilde{\th}}
\def\lt{\tilde{\lambda}}
\def\hl{\hat{\lambda}}
\def\hlt{\hat{\tilde{\lambda}}}
\def\llt{\tilde{l}}
\def\At{\tilde{A}}
\def\Qt{\tilde{Q}}
\def\Rt{\tilde{R}}
\def\Nt{\tilde{N}}

\def\at{\tilde{a}}
\def\st{\tilde{s}}
\def\ft{\tilde{f}}
\def\pt{\tilde{p}}
\def\qt{\tilde{q}}
\def\vt{\tilde{v}}
\def\nt{\tilde{n}}

\def\delb{\bar{\partial}}
\def\bz{\bar{z}}
\def\bD{\bar{D}}
\def\bB{\bar{B}}

\def\bk{{\bf k}}
\def\bl{{\bf l}}
\def\bp{{\bf p}}
\def\bq{{\bf q}}
\def\br{{\bf r}}
\def\bx{{\bf x}}
\def\by{{\bf y}}
\def\bR{{\bf R}}
\def\bV{{\bf V}}

\def\d{\delta}\def\D{\Delta}\def\ddt{\dot\delta}
\def\pa{\partial} \def\del{\partial}
\def\xx{\times}
\def\uno{\mbox{1 \kern-.59em {\rm l}}}
\def\trp{^{\top}}
\def\inv{^{-1}}
\def\dag{{^{\dagger}}}
\def\pr{^{\prime}}
\def\lan{\langle}
\def\ran{\rangle}
\def\rar{\rightarrow}
\def\lar{\leftarrow}
\def\lrar{\leftrightarrow}
\newcommand{\0}{\,\!}      
\def\one{1\!\!1\,\,}
\def\im{\imath}
\def\jm{\jmath}
\newcommand{\tr}{\mbox{tr}}
\newcommand{\slsh}[1]{/ \!\!\!\! #1}
\def\vac{|0\rangle}
\def\lvac{\langle 0|}
\def\hlf{\frac{1}{2}}
\def\ove#1{\frac{1}{#1}}
\def\Box{\square}
\def\ZZ{\mathbb{Z}}
\def\CC#1{({\bf #1})}
\def\bcomment#1{}
\def\bfhat#1{{\bf \hat{#1}}}
\def\VEV#1{\left\langle #1\right\rangle}
\newcommand{\ex}[1]{{\rm e}^{#1}} \def\ii{{\rm i}}
\def\rr{{\rm r}} \def\rs{{\rm s}}\def\rv{{\rm v}}
\def\ri{{\rm i}}\def\rj{{\rm j}}
\newcommand{\lrbrk}[1]{\left(#1\right)}
\newcommand{\sfrac}[2]{{\textstyle\frac{#1}{#2}}}

\def\Li{{\rm Li}_2}

\font\mybb=msbm10 at 12pt
\def\bb#1{\hbox{\mybb#1}}

\font\myBB=msbm10 at 18pt
\def\BB#1{\hbox{\myBB#1}}

\begin{document}

\begin{flushright}
QMUL-PH-10-15
\end{flushright}

\vspace{20pt}

\begin{center}

{\Large \bf  A Note on Dual MHV Diagrams in $\cN=4$ SYM   }
\vspace{11pt}
\vspace{32pt}

{\mbox {\bf Andreas Brandhuber, Bill Spence, Gabriele Travaglini and Gang Yang}}%
\footnote{ {\sffamily \{\tt a.brandhuber, w.j.spence, g.travaglini,
g.yang\}@qmul.ac.uk }}

\bigskip

{\em Centre for Research in String Theory\\
Department of Physics\\
Queen Mary University of London\\
Mile End Road, London, E1 4NS\\
United Kingdom
 }

\bigskip

\vspace{30pt} {\bf Abstract}

\end{center}

\noindent Recently a reformulation of the MHV diagram method in
${\cal N}=4$ supersymmetric Yang-Mills theory in momentum twistor
space was presented and was shown to be equivalent to the
perturbative expansion of the expectation value of a supersymmetric
Wilson loop in momentum twistor space. In this note we present
related explicit Feynman rules in dual momentum space, which should
have the interpretation of Wilson loop diagrams in dual momentum
space. We show that these novel rules are completely equivalent to
ordinary spacetime MHV rules and can be naturally viewed as their
graph dual representation.

\setcounter{page}{0}
\thispagestyle{empty}
\newpage


\setcounter{footnote}{0}

 \section{Introduction}

Witten's seminal work \cite{Witten:2003nn} identified novel twistor space structures previously hidden in scattering amplitudes, linking these to a twistor string theory formulation of ${\cal N}=4$  super Yang-Mills (SYM) theory.
Subsequent work  \cite{Cachazo:2004kj} showed that tree-level  amplitudes could be derived perturbatively from
a novel diagrammatic expansion obtained by joining MHV amplitudes (appropriately continued off shell) together with scalar propagators, based on the geometric twistor picture of amplitudes as sets of intersecting lines.
It was then demonstrated in \cite{Brandhuber:2004yw, Brandhuber:2005kd}  that the  MHV diagram approach extends to the quantum level, by showing that it gives the correct one-loop MHV amplitudes in $\cN=4$ SYM.
Quantum MHV diagrams were later successfully applied to reproducing one-loop MHV amplitudes in $\cN=1$ SYM \cite{Bedford:2004py, Quigley:2004pw}, and to deriving the cut-constructible part of the infinite sequence of one-loop MHV amplitudes in pure Yang-Mills theory \cite{Bedford:2004nh}.
A general proof of the equivalence of MHV diagrams to conventional Feynman rules was later presented in
\cite{Brandhuber:2005kd} for generic one-loop amplitudes in SYM, and for the cut-constructible part of Yang-Mills amplitudes.
It was shown in \cite{Gorsky:2005sf, Mansfield:2005yd} that MHV rules can be obtained from a special change of variables performed on the lightcone Yang-Mills action. In \cite{Boels:2006ir, Boels:2007qn} twistor space actions of gauge theories were constructed and the MHV rules derived from a particular gauge fixing of these actions.

A rather different, dual, prescription to calculate ${\cal N}=4$
amplitudes was found in \cite{am} based on the AdS/CFT correspondence,
where the calculation of amplitudes was mapped to that of a Wilson loop with a lightlike polygonal contour. Somewhat surprisingly, this was found to yield correct results also in the weakly-coupled theory at one loop \cite{Drummond:2007aua, Brandhuber:2007yx},
as well as two loops \cite{Drummond:2007cf,   Drummond:2007au,   Bern:2008ap, Drummond:2008aq}.
An important difference between the Wilson loop approach and unitarity-based methods is that they lead to rather different classes of integral functions. Furthermore, the number of different integral topologies stops growing at a different number of external particles $n$. For example, at two loops there are no new integrals after nine points \cite{Anastasiou:2009kna} while for amplitudes this number is twelve \cite{Vergu:2009tu, ArkaniHamed:2010kv}.
However, so far this Wilson loop approach is limited to MHV amplitudes, and a generalisation to non-MHV amplitudes is not known.%
\footnote{See the very recent papers \cite{Mason:2010yk,sch} for
interesting proposals to overcome this limitation.}

More recently, interesting developments have emerged from the discovery of the important role played by dual superconformal symmetry in the study of the ${\cal N}=4$  theory \cite{Drummond:2006rz,  Drummond:2008vq}. It is then natural to study the role played by dual, or momentum twistors \cite{Hodges:2009hk}, which  led to elegant reformulations of the ${\cal N}=4$ theory amplitudes in momentum twistor space
(see \cite{ArkaniHamed:2010kv} and references therein, for example).
This work has led to a recent proposal
\cite{Bullimore:2010pj} that the MHV diagram method can be formulated in momentum twistor space, with the momentum twistor MHV rules generating compact expressions for integrands of loop amplitudes, which have   manifest dual superconformal symmetry -- up to the  choice of a reference twistor.
Connected with this, in \cite{Mason:2010yk} it was proposed that the expectation value of supersymmetric Wilson loops in momentum twistor space generate all planar amplitudes in the ${\cal N}=4$ theory.
This work suggests the existence of a formulation  of a Wilson loop approach to all ${\rm N}^k$MHV amplitudes directly in dual momentum space.

Ultimately the goal would be to find a derivation of the
amplitude/Wilson-loop duality \cite{am,Drummond:2007aua,
Brandhuber:2007yx} and an extension to all amplitudes in $\cN=4$
SYM. As mentioned above, the two sides of this correspondence lead to rather different
integral representations. What we present in this paper is a set of diagrammatic
rules directly in dual momentum space inspired by the momentum
twistor Wilson loop formulation of Mason and Skinner
\cite{Mason:2010yk}. The answers obtained by these rules are
completely equivalent to the results obtained by supersymmetric  MHV rules (as used in \cite{gg1,gg2, Brandhuber:2004yw}),
divided by the MHV tree-level superamplitude. This
suggests the existence of a dual momentum space formulation of the
Wilson loop approach to all ${\rm N}^k$MHV amplitudes.

The rest of the paper is  organised as follows.
In Section 2 we introduce our diagrammatic rules, while in Section 3 we demonstrate full equivalence with MHV diagrams at tree level. In Sections 4 and 5 we consider one-loop and higher-loop diagrams and find agreement with known results. In Section 6 we conclude with a brief discussion of our results and future research directions.

\section{Feynman rules}
\label{Feynmanrules}

\begin{figure}[h]
\centerline{\includegraphics[height=3cm]{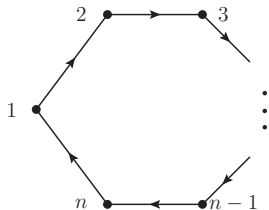} }
\caption{\it Null polygon in dual momentum superspace.}
\label{polygon}
\end{figure}
Here we present the basic Feynman rules for the proposed Wilson loop
formulation of ${\cal N}=4$  amplitudes. We first  consider a
polygonal configuration given by $n$ points $(x_i, \theta_i)$ in
${\cal N}=4$ dual superspace \cite{Drummond:2008vq}, as shown in
Figure \ref{polygon}. Adjacent points $i$ and $i+1$ are null
separated, and  we introduce dual momentum superspace coordinates,
\bea x_i - x_{i+1} = p_i = \lambda_i \tilde \lambda_i ~, \qquad
\theta_i - \theta_{i+1} = \lambda_i \eta_i ~, \eea
with the conventions
\bea x_{ab} = x_a - x_b ~, \qquad \theta_{ab} = \theta_a - \theta_b
~. \eea
We define the spinor $|\ell_{ab}\rangle_\alpha$ as
\bea | \ell_{ab} \rangle \equiv |x_{ab}|\iota ] ~, \eea
where $|\iota]^{\dot \alpha}$ is an arbitrary reference spinor, as used in the MHV rules \cite{Cachazo:2004kj}.

\begin{figure}[h]
\centerline{\includegraphics[height=6.7cm]{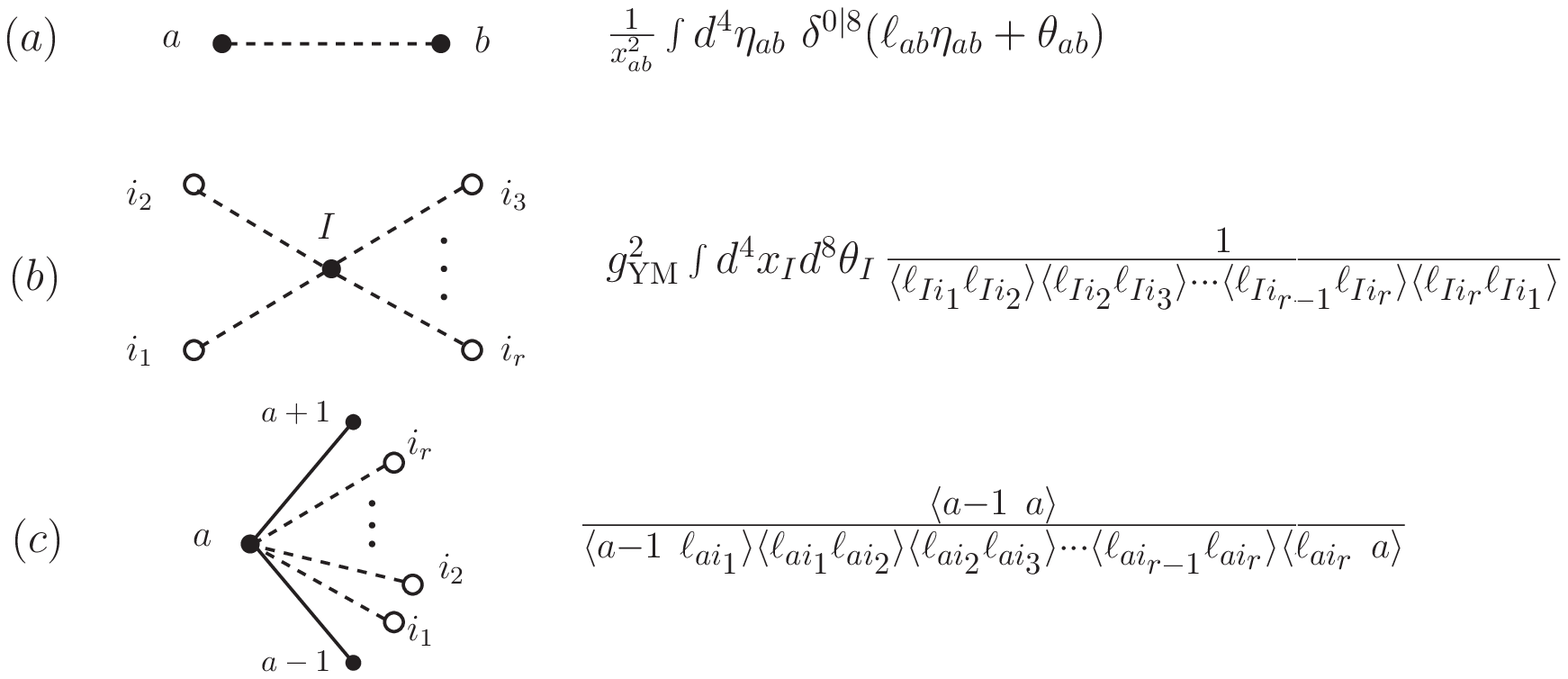} }
\caption{\it Feynman rules. (a) Propagator. (b) Internal vertices.
(c) External vertices.} \label{rules}
\end{figure}

In Figure \ref{rules} we have summarized our diagrammatic rules in
dual superspace or dual MHV rules for the calculation of tree and
loop-level superamplitudes in $\cN=4$ SYM. Note that the result is
the superamplitude divided by the MHV tree-level superamplitude. We
will illustrate these rules in detail in examples presented in later
sections and will give here only some general comments.

The propagator Figure \ref{rules}(a) connects any pair of white dots from internal vertices, depicted in Figure \ref{rules}(b), or external vertices, shown in Figure \ref{rules}(c).
A calculation of an $L$-loop superamplitude ${\cal A}_{{\rm N}^k {\rm
MHV}}^{L\textrm{-loop}}$ would involve $k+L+1$ conventional spacetime MHV vertices.
In the dual momentum picture,
each internal vertex contributes a factor $g_{\rm YM}^2$ and fermionic
degree $-8$, while each propagator contributes fermionic degree 4.
Hence, in order to calculate  ${\cal A}_{{\rm N}^k {\rm
MHV}}^{L\textrm{-loop}}$, we need to sum over all possible diagrams
with
\bea \#({\rm internal \ vertices}) = L ~, \qquad \#({\rm propagators}) = k + 2
L ~. \eea

Before we  move on to concrete tree and loop-level examples, we would like to make some general comments on the possible origin of these rules. The propagator resembles a standard scalar propagator, except for the presence of a fermionic integration. The internal vertices   are most naturally interpreted as
insertions arising from an MHV-type Lagrangian in dual momentum space, and finally,
we expect the external vertices to arise from the expansion of an appropriate Wilson loop.
In this respect we notice that two interesting proposals for Wilson loops in dual momentum space have recently appeared \cite{Mason:2010yk,sch}, and it would be natural to expect that our rules are closely related to these.
Finally, we note that our diagrammatic approach is, at least geometrically, a direct translation of the momentum twistor Wilson loop approach of \cite{Mason:2010yk}, which makes the relation to supersymmetric MHV rules manifest.

\section{Tree amplitudes}

We begin by  considering diagrams without internal vertices. As we shall see below, these turn out to generate the tree amplitudes of the $\cN=4$ theory, divided by the tree MHV amplitude.\\

\noindent
{\bf MHV tree amplitudes}

In this case the number of propagators and internal vertices is zero.
It can be viewed as a limiting case of the external vertex in Figure \ref{rules}(c)
with all dashed legs removed, to which we can assign the value $1$. This
matches the MHV tree amplitude divided by itself.\\

\noindent
{\bf NMHV tree amplitudes}

\begin{figure}[h]
\centerline{\includegraphics[height=3.5cm]{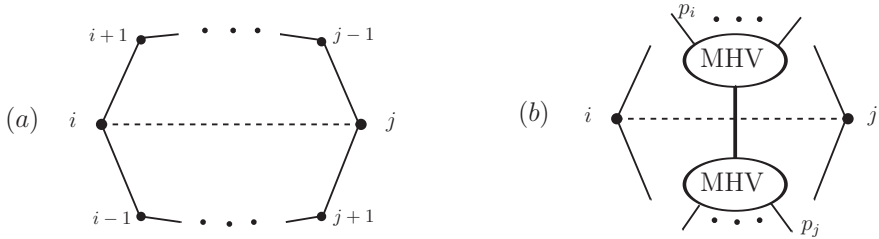} }
\caption{\it (a) Dual MHV diagrams contributing to NMHV tree amplitudes. (b) Here we show the conventional  MHV diagrams together with the dual diagrams.} \label{NMHV}
\end{figure}

The next simplest diagram without internal vertices is one with a single propagator linking two external vertices, labelled by $i$ and $j$ in Figure \ref{NMHV}(a) below.

Applying our rules we obtain the expression
\bea {\langle i\!-\!1 ~i \rangle \over \langle i\!-\!1 ~\ell_{ij}
\rangle \langle \ell_{ij} ~i \rangle} {\langle j\!-\!1 ~j \rangle
\over \langle j\!-\!1 ~\ell_{ij} \rangle \langle \ell_{ij} ~j
\rangle} {1\over x_{ij}^2} \int d^4 \eta_{ij} ~ \delta^{0|8}
(\ell_{ij} \eta_{ij} + \theta_{ij}) ~. \eea
This is precisely the expression of the corresponding MHV diagram
depicted in Figure \ref{NMHV}(b), where we also show the dual momenta $x_i$ and $x_j$
in order to illustrate the graph duality between the two diagrams. Note that
this term above is nothing but  the superconformally invariant
$R$-function \cite{Drummond:2008vq} $R_{*;ij} = [*,i\!-\!1,i,j\!-\!1,j]$ in the notation of \cite{Mason:2009qx,Bullimore:2010pj} with reference twistor $Z_*=(0,\iota,0)$.\\

\noindent
{\bf $\textrm{N}^2$MHV tree amplitudes}

\begin{figure}[h]
\centerline{\includegraphics[height=3.5cm]{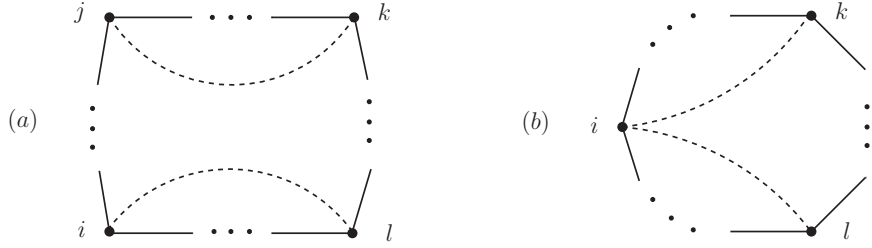} }
\caption{\it Dual MHV diagrams contributing to $\textrm{N}^2$MHV tree amplitudes. The corresponding ordinary MHV  diagrams are depicted in Figure \ref{N2MHV-mhvdiag} below. } \label{N2MHV}
\end{figure}

The next simplest dual MHV diagrams without internal vertices are depicted in Figure \ref{N2MHV}.
\begin{figure}[h]
\centerline{\includegraphics[height=1.65cm]{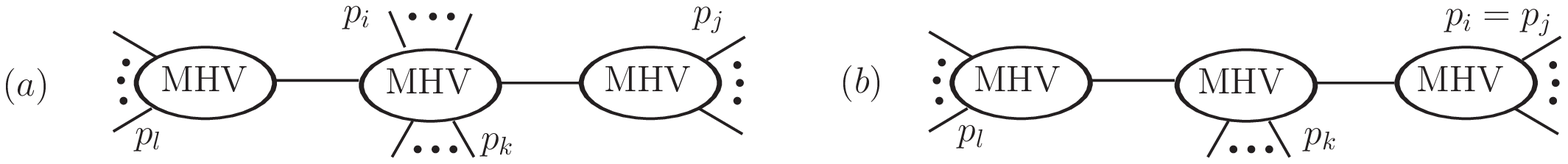} }
\caption{\it Conventional  MHV diagrams for   $\textrm{N}^2$MHV tree amplitudes.} \label{N2MHV-mhvdiag}
\end{figure}
These two kinds of diagrams will turn out to generate the ${\rm N}^2$MHV tree amplitude.
The first one, given in Figure \ref{N2MHV}(a), gives
\bea && {\langle i\!-\!1 ~i \rangle \over \langle i\!-\!1 ~\ell_{il}
\rangle \langle \ell_{il} ~i \rangle} {\langle l\!-\!1 ~l \rangle
\over \langle l\!-\!1 ~\ell_{il} \rangle \langle \ell_{il} ~l
\rangle} {1\over x_{il}^2} \int d^4 \eta_{il} ~ \delta^{0|8}
(\ell_{il} \eta_{il} + \theta_{il}) \nonumber\\ && \times {\langle
j\!-\!1 ~j \rangle \over \langle j\!-\!1 ~\ell_{jk} \rangle \langle
\ell_{jk} ~j \rangle} {\langle k\!-\!1 ~k \rangle \over \langle
k\!-\!1 ~\ell_{jk} \rangle \langle \ell_{jk} ~k \rangle} {1\over
x_{jk}^2} \int d^4 \eta_{jk} ~ \delta^{0|8} (\ell_{jk} \eta_{jk} +
\theta_{jk}) . \eea
The two lines in the expression above are independent as Figure
\ref{N2MHV}(a) shows. This reproduces the result of conventional MHV
diagram in Figure \ref{N2MHV-mhvdiag}(a), divided by tree amplitude.
The result is simply the product of two $R$-functions,
 \bea
 [ *, j\!-\!1, j, k\!-\!1, k ] \  [ *, l\!-\!1, l, i\!-\!1, i ] .
\eea
The second type of diagrams contributing to the tree ${\rm N}^2$MHV
amplitude is given in Figure \ref{N2MHV}(b), which corresponds to
the special case $i=j$ of Figure \ref{N2MHV}(a). This gives
\bea \label{nnmhvtree} && {\langle i\!-\!1 ~i \rangle \over \langle
i\!-\!1 ~\ell_{il} \rangle \langle \ell_{il} ~\ell_{ik} \rangle
\langle \ell_{ik} ~i \rangle} {\langle k\!-\!1 ~k \rangle \over
\langle k\!-\!1 ~\ell_{ik} \rangle \langle \ell_{ik} ~k \rangle}
{\langle l\!-\!1 ~l \rangle \over \langle l\!-\!1 ~\ell_{il} \rangle
\langle \ell_{il} ~l \rangle}\nonumber\\ && \times {1\over x_{il}^2}
\int d^4 \eta_{il} ~ \delta^{0|8} (\ell_{il} \eta_{il} +
\theta_{il}) {1\over x_{ik}^2} \int d^4 \eta_{ik} ~ \delta^{0|8}
(\ell_{ik} \eta_{ik} + \theta_{ik})\ .  \eea
It is readily seen that this is the same as the expression one obtains from the corresponding MHV diagram, in   Figure \ref{N2MHV-mhvdiag}(b).  In the notation of  \cite{Bullimore:2010pj}, this diagram gives
\bea
\label{thisone}
 [ *, i\!-\!1, i, k\!-\!1, k ] \  [ *, \widehat{k\!-\!1}, k, l\!-\!1, l ] ,
\eea
where the hat refers to the shifting of the momentum twistor $ \widehat{k\!-\!1} =
(k\!-\!1,k)\cap (*, i\!-\!1,i)$.
It is shown in  \cite{Bullimore:2010pj} that the twistor
expression \eqref{thisone} for this tree ${\rm N}^2$MHV amplitude matches that
obtained from the MHV diagram approach, although this comparison is somewhat more involved than
the one here.

Note that the multi-leg vertices (Figure 2(c) with $r\!>\!1$) make
an appearance here with a vertex with $r\!=\!2$ being used, and that
these vertices automatically take  account of the special case that
in the twistor picture requires the somewhat mysterious introduction
of  hatted twistor rules. We expect this to be a generic feature,
i.e.~that the \lq\lq shifted twistors\rq\rq\ of the rules of
\cite{Bullimore:2010pj} will in general be automatic consequences of
the Feynman rules proposed here.

\begin{figure}[b]
\begin{center}
\scalebox{0.70}{
\includegraphics[width=8cm]{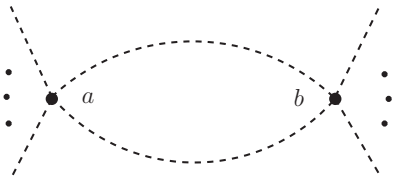}
}
\end{center}
\caption{\it A dual MHV diagram potentially contributing to
$\textrm{N}^2$MHV amplitudes which is not present in the
conventional MHV diagram expansion. Its explicit contribution
vanishes. } \label{vanish}
\end{figure}

Finally, we notice that there is a special class of dual MHV
diagrams that apparently   contributes to tree-level  ${\rm N}^2$MHV
superamplitudes, that is obtained by joining two external vertices with $r=2$ with
two propagators, such as taking both $i=j$ and $k=l$ in Figure
\ref{N2MHV}(a). Such a diagram, depicted in Figure \ref{vanish},
corresponds to a conventional MHV diagram with three vertices  where
the middle one would be bivalent, with its two legs connected to
the adjacent MHV vertices. Due to the absence of conventional
two-point MHV vertices, we expect such diagrams to vanish in the
dual formulation, as we now check.

Denoting by  $x_a$ and $x_b$ the positions of the external vertices in dual momentum space, we find this diagram to be equal to
\bea \label{vanam} && {\lan a\!-\!1 ~a\ran \over \lan a\!-\!1
~l_{ab} \ran \lan l_{ab} ~l_{ab} \ran \lan l_{ab} ~a \ran} \, {\lan
b\!-\!1 ~b\ran \over \lan b\!-\!1 ~l_{ab} \ran \lan l_{ab} ~l_{ab}
\ran \lan l_{ab} ~b \ran} \,
\nonumber \\
&&\int\!d^4\eta_{ab} \,
{\delta^{0|8} (\eta_{ab} l_{ab} + \theta_{ab}) \over x_{ab}^2 }
\int\!d^4\eta_{ab}^\prime \,
{\delta^{0|8} (\eta_{ab}^\prime l_{ab} + \theta_{ab}) \over x_{ab}^2 } \ .
\eea
The prefactor in \eqref{vanam} contains a double pole $1/\lan l_{ab} l_{ab} \ran^2$. However, one has in general
\be
\int\!d^4\eta d^4\eta^\prime \,
\delta^{0|8} (\eta l+ \theta)
\delta^{0|8} (\eta^\prime l^\prime + \theta ) \ = \ {1\over 16} \lan l l^\prime \ran^4 \prod_{A=1}^4
\lan \theta^A \theta^A \ran
 \ ,
\ee and therefore the second line of \eqref{vanam} provides  a
factor of  $\lan l_{ab} l_{ab}\ran^4$. This  compensates the double
pole from the prefactor, and makes this class of diagrams
vanish.\\

\noindent
{\bf Generic tree amplitudes}

It is clear from the examples discussed above that there is a direct
correspondence between dual MHV diagrams without internal
vertices, and tree-level MHV diagrams contributing to generic
${\rm N^{\it k}}${\rm MHV} amplitudes. This also
covers all special cases where dual momenta coincide (c.f.~Figure
\ref{N2MHV}(b) above). This is the same diagrammatic duality
discussed in Section 4.3 of \cite{Bullimore:2010pj}, in the context of
momentum twistor space.


\section{One-loop amplitudes}

In this section we start to consider loop amplitudes which requires us
to include also internal vertices, represented in Figure 2(b), in the dual MHV diagrams. At one loop we thus have to consider dual MHV diagrams with one internal vertex.\\

\noindent
{\bf One loop MHV amplitudes}

The simplest Wilson loop with one internal vertex is given in Figure
\ref{MHV-1-loop} below. Note that the dual superspace position
$(x_I,\theta_I)$ of the internal vertex has to be integrated over.
This diagram is equal to
\bea &&  g_{\rm YM}^2 \int d^4 x_I d^8 \theta_I {1\over \langle
\ell_{iI} \ell_{Ij}
\rangle \langle \ell_{Ij} \ell_{iI} \rangle } \ {\langle i\!-\!1 ~i \rangle \over \langle i\!-\!1 ~\ell_{iI}
\rangle \langle \ell_{iI} ~i \rangle} {\langle j\!-\!1 ~j \rangle
\over \langle j\!-\!1 ~\ell_{Ij} \rangle \langle \ell_{Ij} ~j
\rangle}  \nonumber\\
&& {1\over x_{iI}^2} \int d^4 \eta_{iI} ~ \delta^{0|8} (\ell_{iI}
\eta_{iI} + \theta_{iI}) {1\over x_{Ij}^2} \int d^4 \eta_{Ij} ~
\delta^{0|8} (\ell_{Ij} \eta_{Ij} + \theta_{Ij}) ~, \eea
where the integration over the fermionic variables gives the factor
$\langle \ell_{iI} \ell_{Ij} \rangle^4$. This yields exactly the
result obtained from the one-loop MHV diagram in Figure
\ref{MHV-1-loop}(b) (divided by the tree amplitude).
\begin{figure}[h]
\centerline{\includegraphics[height=3.5cm]{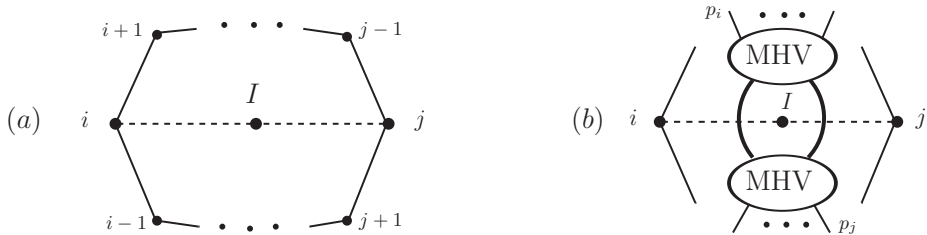} }
\caption{\it One-loop MHV  amplitudes.} \label{MHV-1-loop}
\end{figure}

\noindent
{\bf Generic one-loop amplitudes}

Rather than reviewing the next simplest case, the one-loop NMHV
amplitude, we consider instead the general structure of one-loop
amplitudes in this formulation. The corresponding general structure
 in momentum twistor space  has been discussed in
Section 5.4 of \cite{Bullimore:2010pj}. By virtue of the tree-level
results already obtained above,  one may restrict the attention to
diagrams containing one internal $n$-point vertex at $x_I$, whose
legs connect to $m$ distinct points $x_{j_r}$ ($r=1,...,m$).%
\footnote{ It is
easy to see that if more than one leg connect to the same point, the
result is zero.}
 The vertex generates a factor
\bea g_{\rm YM}^2 \int d^4 x_I d^8 \theta_I {1\over \langle
\ell_{Ij_1} \ell_{Ij_2} \rangle \langle \ell_{Ij_2} \ell_{Ij_3}
\rangle \cdots \langle \ell_{I j_{m\!-\!1}} \ell_{I j_m} \rangle
\langle \ell_{I j_m} \ell_{I j_1} \rangle } ~. \eea
After performing the $\eta$
integrations coming from the propagators,
each propagator connecting the internal vertex to the point
$x_{j_r}$ generates a term,
\bea {1\over x_{Ij_r}^2}  \delta^{0|4} ( \langle \ell_{Ij_r}
|\theta_{Ij_r}\rangle) ~. \eea
Finally, each external point $x_{j_r}$ with a propagator linked to
it generates a term
\bea { \langle j_r\!-\!1\ j_r \rangle \over \langle j_r\!-\!1\
\ell_{Ij_r} \rangle \langle \ell_{Ij_r}\  j_r  \rangle }.
 \eea
Combining these three terms for all legs of the internal vertex one obtains the expression
\bea
\label{1loopgeneral}
    g_{\rm YM}^2\prod_{r=1}^m \frac{\langle j_r\!-\!1\, j_r\rangle \  \delta^{0|4} (\langle
    \ell_{Ij_r}|
\theta_{Ij_r} \rangle) }
    {x_{Ij_r}^2\, \langle j_r\!-\!1\,\ell_{Ij_r}\rangle\,\langle\ell_{Ij_r}\,j_r\rangle\,\langle\ell_{Ij_r}\,\ell_{Ij_{r\!+\!1}}\rangle}\, .
\eea
This is precisely the expression (5.22) of  \cite{Bullimore:2010pj}, obtained from the usual MHV rules and corresponds directly to the product of $R$ invariants that arises from applying the momentum twistor MHV rules discussed there. By this argument we see that the dual MHV rules proposed here correctly generate generic one-loop amplitudes as formulated by standard MHV diagrams.

\section{Two loops}

\begin{figure}[h]
\centerline{\includegraphics[height=2.7cm]{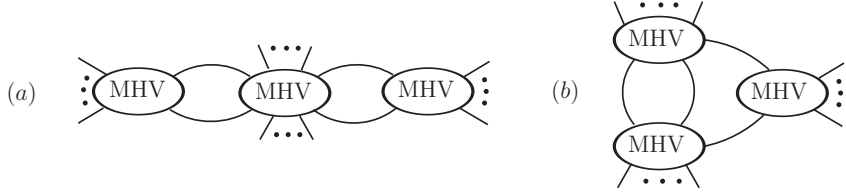} }
\caption{\it Two-loop MHV diagrams. (a) is the double-bubble, and
(b) is the bubble-triangle.} \label{2loop-mhv}
\end{figure}

In this section we construct the integrand of the two-loop MHV
amplitude from dual MHV rules. We will see that this integrand is
the same as the standard integrand produced using spacetime MHV
rules after dividing by the tree-level MHV amplitude. There
are two main topologies of diagrams to consider, namely the
double-bubble and the bubble-triangle topology, as shown in
Figure \ref{2loop-mhv}. Special cases of such topologies arise when
some of the external dual momenta coalesce.
\begin{figure}[h]
\centerline{\includegraphics[height=6cm]{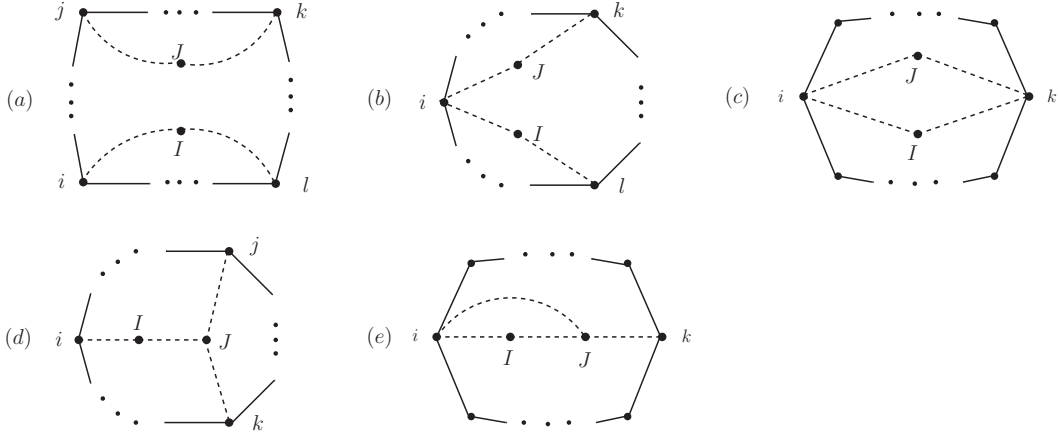} }
\caption{\it Two-loop MHV amplitudes. These graphs have two internal vertices
and four propagators. Diagrams in the first line correspond to
double-bubble MHV diagrams, and diagrams in the second line
correspond to the bubble-triangle MHV diagrams.} \label{MHV-2loop}
\end{figure}

In our description, we draw all possible diagrams that
have two internal vertices and four propagators. The non-vanishing diagrams
are shown in Figure \ref{MHV-2loop}. We consider them in turn.

Using the rules described in Section \ref{Feynmanrules}, we arrive
at the following integrand for Figure \ref{MHV-2loop}(a):
\beqa \label{2loopMHV-1} &&g_{\rm YM}^2 \int d^4 x_I d^8 \theta_I
{1\over \langle \ell_{Ii} \ell_{Il} \rangle^2 } \, g_{\rm YM}^2 \int
d^4 x_J d^8 \theta_J {1\over \langle \ell_{Jj} \ell_{Jk} \rangle^2 }
\nonumber \\
&& {1\over x_{iI}^2} \int d^4 \eta_{iI} ~ \delta^{0|8} (\ell_{iI}
\eta_{iI} + \theta_{iI}) {1\over x_{Il}^2} \int d^4 \eta_{Il} ~
\delta^{0|8} (\ell_{Il} \eta_{Il} + \theta_{Il})
\nonumber \\
&& {1\over x_{jJ}^2} \int d^4 \eta_{jJ} ~ \delta^{0|8} (\ell_{jJ}
\eta_{jJ} + \theta_{jJ}) {1\over x_{Jk}^2} \int d^4 \eta_{Jk} ~
\delta^{0|8} (\ell_{Jk} \eta_{Jk} + \theta_{Jk})
\nonumber \\
&&{\lan i\!-\!1\, i \ran\over
 \lan i\!-\!1 \,\ell_{Ii} \ran \lan \ell_{I i}\,i \ran }\,
 {\lan j\!-\!1 \,j\ran\over
 \lan j\!-\!1\, \ell_{jJ} \ran \lan \ell_{jJ}\, j \ran }\,
 {\lan k\!-\!1\, k\ran\over
 \lan k\!-\!1 \,\ell_{Jk} \ran \lan \ell_{J k}\,k \ran }\,
 {\lan l\!-\!1 \,l\ran\over
 \lan l\!-\!1\, \ell_{lI} \ran \lan \ell_{lI}\, l \ran }
 \  .
 \eeqa
The first line in  \eqref{2loopMHV-1} corresponds to the insertion
of two internal two-point vertices, the second and third line to four
propagators connecting the two-point vertices with four external vertices,
appearing in the last line, located at dual points
$x_i$,  $x_j$, and $x_k$,  $x_m$. The fermionic integrations can be
performed immediately, and give $\lan \ell_{iI} \ell_{lI} \ran^4
\lan \ell_{jJ} \ell_{kJ} \ran^4$. After this, \eqref{2loopMHV-1}
is immediately found to coincide with the integrand of the corresponding
spacetime two-loop MHV diagram.

Next we consider the class of diagrams where $x_i = x_j$,
corresponding to Figure \ref{MHV-2loop}(b). These are given by
\beqa \label{2loopMHV-2} && g_{\rm YM}^2 \int d^4 x_I d^8 \theta_I
{1\over \langle \ell_{Ii} \ell_{Il} \rangle^2 } \, g_{\rm YM}^2 \int
d^4 x_J d^8 \theta_J {1\over \langle \ell_{Ji} \ell_{Jk} \rangle^2 }
\nonumber \\
&& {1\over x_{iI}^2} \int d^4 \eta_{iI} ~ \delta^{0|8} (\ell_{iI}
\eta_{iI} + \theta_{iI}) {1\over x_{Il}^2} \int d^4 \eta_{Il} ~
\delta^{0|8} (\ell_{Il} \eta_{Il} + \theta_{Il})
\nonumber \\
&& {1\over x_{iJ}^2} \int d^4 \eta_{iJ} ~ \delta^{0|8} (\ell_{iJ}
\eta_{iJ} + \theta_{iJ}) {1\over x_{Jk}^2} \int d^4 \eta_{Jk} ~
\delta^{0|8} (\ell_{Jk} \eta_{Jk} + \theta_{Jk})
\nonumber \\
&&{\lan i\!-\!1\, i \ran\over
 \lan i\!-\!1 \,\ell_{Ii} \ran \lan \ell_{I i}\,\ell_{iJ} \ran     \lan \ell_{iJ}\, i \ran     }\,
 {\lan k\!-\!1 \,k\ran\over
 \lan k\!-\!1\, \ell_{kJ} \ran \lan \ell_{kJ}\,k \ran }\,
 {\lan l\!-\!1 \,l\ran\over
 \lan l\!-\!1\, \ell_{lI} \ran \lan \ell_{lI}\, l \ran }
 \  .
 \eeqa
Compared to \eqref{2loopMHV-1}, we notice the appearance  in the fourth line of \eqref{2loopMHV-2} of an external vertex with $r=2$ -- see Figure \ref{rules}(c).
Once again fermionic integrations are trivial to perform and give
$\lan \ell_{iI} \ell_{lI} \ran^4 \lan \ell_{iJ} \ell_{kJ} \ran^4$,
and one then recognises instantly that \eqref{2loopMHV-2} is equal
to the corresponding spacetime MHV diagram.

There is an additional case with the same double-bubble topology,
where the middle MHV vertex has no external legs -- it is only
attached to two adjacent MHV vertices forming the two loops of the
diagram. This correspond to Figure \ref{MHV-2loop}(c), and is also
immediately seen to reproduce the corresponding spacetime MHV
result.

The last diagram topology to consider is the triangle-bubble. In
the most general of diagrams, one has an internal three-point and an internal two-point
vertex, connected together by propagators and connected to external vertices
given in Figure \ref{rules}(c) (with $r\!=\!1$, in the notation of
the Figure).  The diagram is depicted in Figure \ref{MHV-2loop}(d) and equals
\beqa \label{2loopMHV-3} &&
g_{\rm YM}^2 \int d^4 x_I d^8 \theta_I {1\over \langle \ell_{Ii}
\ell_{IJ} \rangle^2 } \, g_{\rm YM}^2 \int d^4 x_J d^8 \theta_J
{1\over \langle \ell_{IJ} \ell_{Jj} \rangle \langle \ell_{Jj} \ell_{Jk}
\rangle \ell_{Jk} \ell_{IJ} \rangle }
\nonumber \\
&& {1\over x_{iI}^2} \int d^4 \eta_{iI} ~ \delta^{0|8} (\ell_{iI}
\eta_{iI} + \theta_{iI})
{1\over x_{IJ}^2}
\int d^4 \eta_{IJ} ~
\delta^{0|8} (\ell_{IJ} \eta_{IJ} + \theta_{IJ})
\nonumber \\
&&
{1\over x_{Jj}^2} \int
d^4 \eta_{Jj} ~
\delta^{0|8} (\ell_{Jj} \eta_{Jj} + \theta_{Jj})
{1\over x_{kJ}^2} \int d^4 \eta_{kJ} ~ \delta^{0|8} (\ell_{kJ}
\eta_{kJ} + \theta_{kJ})
\nonumber \\
&& {\lan i\!-\!1\, i \ran\over
 \lan i\!-\!1 \,\ell_{Ii} \ran \lan \ell_{I i}\,  i \ran
   }\,
 {\lan j\!-\!1 \,j\ran\over
 \lan j\!-\!1\, \ell_{jJ} \ran \lan \ell_{jJ}\,j \ran }\,
 {\lan k\!-\!1 \,k\ran\over
 \lan k\!-\!1\, \ell_{kJ} \ran \lan \ell_{kJ}\, k \ran }
 \  .
 \eeqa
A special case occurs when $i=j$ as shown in
Figure \ref{MHV-2loop}(e), which is also immediately seen to
reproduce the corresponding spacetime MHV result.
\begin{figure}[h]
\centerline{\includegraphics[height=2.7cm]{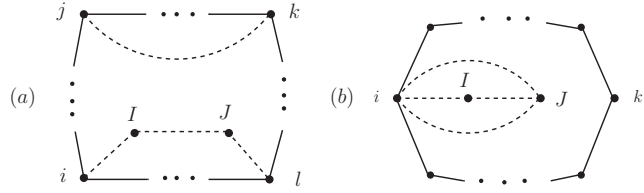} }
\caption{\it These diagrams give zero contribution.}
\label{MHV-2loop-vanish}
\end{figure}
\begin{figure}[h]
\centerline{\includegraphics[height=2.4cm]{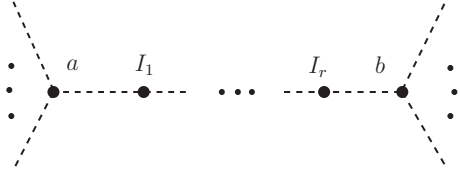} }
\caption{\it
The diagram which contains a chain of two-point
vertices ($r>1$) also gives a vanishing  contribution.}
\label{vanish2}
\end{figure}

There are other graphs,  shown in
Figure \ref{MHV-2loop-vanish}, but they all give  vanishing  contribution. For
example, when there is a structure of a chain of internal two-point vertices,
as in Figure \ref{vanish2}, the result is zero. Indeed, considering only the
fermionic part, we have
\bea && \int d^8 \theta_{I_1} \cdots \theta_{I_r} d^4 \eta_{aI_1}
\cdots  d^4 \eta_{I_r b} ~\delta^{0|8}(\ell_{aI_1}\eta_{aI_1} +
\theta_{aI_1}) \cdots \delta^{0|8}(\ell_{I_rb}\eta_{I_rb} +
\theta_{I_rb}) \nonumber\\ &=& \int d^4 \eta_{aI_1} \cdots  d^4
\eta_{I_r b} ~\delta^{0|8}(\ell_{aI_1}\eta_{aI_1} +\cdots +
\ell_{I_rb}\eta_{I_rb} + \theta_{ab}) ~, \eea
and unless $r=1$ the integration always gives zero. This is what happens
for Figure \ref{MHV-2loop-vanish}(a).
Figure \ref{MHV-2loop-vanish}(b)  contains the structure made by joining
two internal two-point vertices with two propagators, which also
vanishes as discussed in section 3.
%



\section{Discussion}

The proposal given here provides a MHV diagram formulation of all
$\cN=4$ amplitudes in dual momentum space, which is dual to the
conventional MHV diagram approach \cite{Cachazo:2004kj}. This is the
dual momentum version (i.e. dual space-time version) of the momentum
twistor space Wilson loop approach of \cite{Mason:2010yk}. Whilst
one may appeal to that formulation to argue that the Feynman rules
given here must necessarily arise from the $\cN=4$ theory, there is
presumably a direct argument for this. The internal vertices of
Figure \ref{rules}(b) appear to be of MHV origin, and together with
the kinetic terms giving the propagator of Figure \ref{rules}(a) may
be related to the $\cN=4$ formulation of \cite{Gorsky:2005sf,
Mansfield:2005yd}. As discussed in Section \ref{rules}, the external
vertices of Figure \ref{rules}(c) should then arise from the
definition of a  Wilson loop, perhaps along the lines of the twistor
space argument presented in \cite{Mason:2010yk}. It would also be
interesting to relate the rules here to some lightcone limit of
correlation functions, along the lines of recent work
\cite{Alday:2010zy, Eden:2010zz, Eden:2010ce} (see also \cite{sch}).

There are some interesting features of the Feynman rules in Figure
\ref{rules} that may throw light on these. We note that the external
vertices are inserted at points in dual momentum space, with the
simplest  insertion  corresponding to adding one soft function, and
$n$-point insertions corresponding to a multiple soft function with
$n$ legs becoming soft.   Multiplying by soft functions is the
natural way to add particles in the Yangian symmetric
\cite{Drummond:2009fd} formalism of \cite{ArkaniHamed:2010kv,
ArkaniHamed:2009dn}, and this may bear further examination. These
multiple field insertions in the external vertices are relevant when
some dual momentum regions coalesce, and the corresponding rules are
very simple modifications of those for less exceptional cases. Their
origin should come from the expansion of a dual momentum Wilson
loop. The fact that the rules presented here appear to automatically
take account of shifted twistors in \cite{Mason:2010yk} also
supports this.

Finally, we notice that collinear and multi-particle singularities
of the dual MHV diagrams correspond to dual momenta becoming
lightlike separated, and  may  therefore be obtained by a lightcone
operator product expansion. It would also be interesting to relate
this to recent work of \cite{Alday:2010ku}.


 \vspace{0.7cm}
\section*{Acknowledgements}

This work was supported by the STFC Rolling Grant  ST/G000565/1. WJS
was supported by a Leverhulme Research Fellowship, GT by an EPSRC
Advanced Research Fellowship EP/C544242/1, and GY by a STFC
Postdoctoral Fellowship.


\end{document}